# Polarimetric Control of Reflective Metasurfaces

Xavier Artiga, Daniele Bresciani, Hervé Legay, Julien Perruisseau-Carrier, *Member, IEEE*

*Abstract*—This letter addresses the synthesis of reflective cells approaching a given desired Floquet's scattering matrix. This work is motivated by the need to obtain much finer control of reflective metasurfaces by controlling not only their co-polarized reflection but also their cross-coupling behavior. The demonstrated capability will enable more powerful design approaches –involving all field components in phase and magnitude– and consequently better performance in applications involving reflective metasurfaces. We first expose some fundamental theoretical constraints on the cell scattering parameters. Then, a successful procedure for controlling all four scattering parameters by applying parallelogram and trapezoid transformations to square patches is presented, considering both normal and oblique incidence.

## I. INTRODUCTION

METASURFACES are periodic or quasi-periodic planar structures of sub-wavelength elements, which have received significant attention from antenna designers over recent years. For instance, they can be found as reflectors in many applications from microwave to optics, such as reflectarrays and folded-reflectarrays [1], to the similar concept of Generalized Snell's law [2], in quasi-optical systems (FSS and polarizers) [3] or in Fabry-Perot antennas (partially reflective surface) [4]. In these applications the design of the surface generally focuses on synthesizing desired co-polar reflection coefficients but no attempt on controlling the cross-coupling coefficients is made. However, a full polarimetric control could provide new capabilities to such surfaces enabling for example polarization transformations or efficient cross-polarization cancellation techniques [5].

Recently, polarization control of metasurfaces has been demonstrated in the context of leaky-wave antennas [6-7]. However the design approach there is based on guiding surface waves rather than directly controlling the reflection or transmission coefficient of the surface under plane wave excitation, as needed in other applications of metasurfaces such as reflectarrays, FSS, or Fabry-Perot antennas.

In this paper a novel method for controlling the four scattering terms of reflective cells in periodic or quasi periodic metasurfaces is presented. After deriving the theoretical constraints imposed by reciprocity and energy conservation over the scattering matrices of reflective cells, we demonstrate that the application of parallelogram and trapezoid transformations over symmetric square patches provides not only control over the magnitude of the cross-coupling terms but also over their phases. Finally a guideline for the synthesis of a unit cell with a given required S-matrix is developed.

## II. FUNDAMENTAL CONSTRAINTS OF THE SCATTERING MATRIX

### A. Floquet scattering matrix

Synthesis methods for quasi-periodic reflective surfaces assume local periodicity, so that the reflection properties of a cell can be determined by its Floquet's harmonics scattering matrix. For practical cell electrical sizes, only one TE and TM incidence like Floquet's harmonic propagate and the scattering matrix writes:

$$S_{FL} = \begin{bmatrix} s_{11} & s_{12} \\ s_{21} & s_{22} \end{bmatrix} \equiv \begin{bmatrix} \frac{TE_R}{TE_I}\Big|_{TM_I=0} & \frac{TE_R}{TM_I}\Big|_{TE_I=0} \\ \frac{TM_R}{TE_I}\Big|_{TM_I=0} & \frac{TM_R}{TM_I}\Big|_{TE_I=0} \end{bmatrix} \quad (1)$$

### B. Reciprocity

It is well-known that two-port physical devices made of reciprocal material meet the usual 'reciprocity condition' $s_{12}=s_{21}$. However, for the general case of an asymmetric cell under oblique incidence, this constraint does not apply as such to the matrix defined by (1) due to the fact that incident and reflected ports are not the same. Let us consider Fig 1. The incident and reflected elevation angles are the same ($\theta_R=\theta_I$) but the azimuth ones differ from 180° ($\varphi_R=\varphi_I+\pi$) due to the reflection on the surface. Note that in this figure, the orientation of the TE-TM modes is arbitrary, but selected to enforce the usual -1 reflection coefficient for a perfect electric conductor surface.

The mathematical expression of reciprocity in this case can be also deduced from a careful inspection of Fig. 1, where the TE-TM vectors corresponding to $s_{12}(\varphi_i=\varphi_0)$, $s_{21}(\varphi_i=\varphi_0)$, and

Manuscript received September 7, 2012. This work was supported by Thales Alenia Space, France, and by the Swiss National Science Foundation Professorship PP00P2-133583.

X. Artiga is with the Centre Tecnològic de Telecomunicacions de Catalunya (CTTC), Barcelona, Spain (xavier.artiga@cttc.es).

J. Perruisseau-Carrier is with the group for Adaptive MicroNano Wave Systems, LEMA/Nanolab, École Polytechnique Fédérale de Lausanne (EPFL), Lausanne, Switzerland (julien.perruisseau-carrier@epfl.ch).

D. Bresciani and H. Legay are with Thales Alenia Space, Toulouse, France ({daniele.bresciani, herve.legay}@thalesaleniaspace.com).

$s_{21}(\varphi_i=\varphi_0+\pi)$ are shown [sub-figures (a), (b) and (c), respectively]. It is observed that the reflection of a TE wave from a TM incident one [Fig. 1(a)] is not reciprocal to the reflection of a TM wave from a TE incident one [Fig. 1(b)], mathematically

$$s_{12}(\varphi_i = \varphi_0) \neq s_{21}(\varphi_i = \varphi_0) \quad (2)$$

In contrast, the situation of Fig. 1(c) is reciprocal to that of Fig. 1(a). In other words, when TE and TM incident and reflected are interchanged, the angle of incidence φ must be shifted 180° to represent the reciprocal situation. Hence for the matrix defined in (1), reciprocity is in general expressed as

$$s_{12}(\varphi_i = \varphi_0) = s_{21}(\varphi_i = \varphi_0 + \pi) \quad (3)$$

It is worth noticing that in (2) only the phases of both sides of the inequality are different, since by energy conservation (see Section II.C) $|s_{12}(\varphi_i)| = |s_{21}(\varphi_i)|$.

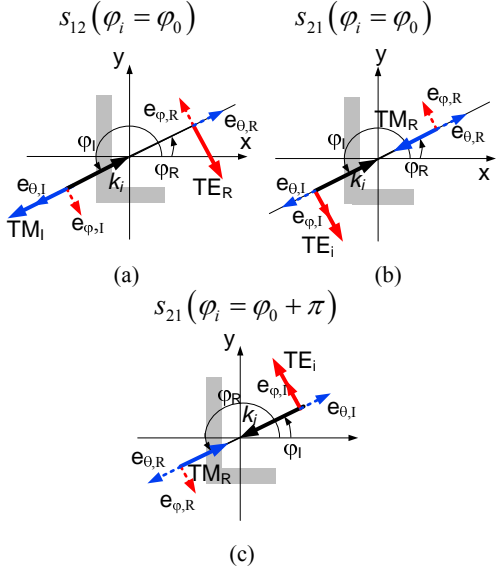

Fig. 1. TE-TM unitary vectors for the calculation of: a) $s_{12}(\varphi_i=\varphi_0)$; b) $s_{21}(\varphi_i=\varphi_0)$; c) $s_{21}(\varphi_i=\varphi_0+\pi)$. The reciprocal situation of case a) is not case b), but case c)

### C. Energy conservation

It is important to recall here energy conservation expressions since, together with the particular reciprocity condition (3), they set the absolute limitation on the synthesizable cell scattering matrices targeted in this work:

$$|s_{11}|^2 + |s_{21}|^2 = 1 \quad (4)$$

$$|s_{22}|^2 + |s_{12}|^2 = 1 \quad (5)$$

$$s_{11}^* s_{12} + s_{21}^* s_{22} = 0 \quad (6)$$

Now, developing (6) in terms of magnitude and phase and substituting in (4) and (5) yields

$$|s_{11}| = |s_{22}| \quad (7)$$

$$|s_{12}| = |s_{21}| \quad (8)$$

$$\angle s_{11} + \angle s_{22} = \angle s_{12} + \angle s_{21} \pm (2n+1)\pi \quad (9)$$

This last expression provides an interesting relation between the phases of the four scattering parameters. Obviously equations (3-5) and (7-9) have to be carefully considered during the design process since they show that a unit cell cannot be synthesized to implement any given scattering matrix target with arbitrary precision, as further discussed next.

## III. CONTROL OF THE REFLECTIVE CELL SCATTERING MATRIX

The control of the four scattering terms of a reflective cell can be achieved as follows. First, the desired co-polar phases ($s_{11}$, $s_{12}$) are obtained by selecting the size of a square patch [1]. Second, parallelogram and trapezoid transformations are applied to the patch in order to achieve the desired cross-coupling coefficients ($s_{12}$, $s_{21}$), in such a manner to have minimal effect on the co-term, as explained next.

### A. Parallelogram and trapezoid transformations

The parallelogram transformation is used to provide control of the cross-coupling terms magnitude, and consists in tilting two opposite sides of the square patch by the same angle. As depicted in Fig. 2, the sides perpendicular to the x-axis are tilted in such a way to preserve the global patch size so that co-polar reflection phases are not significantly affected. Though the cross-coupling magnitude control is limited by (4), in most applications the required cross-coupling magnitudes will be much smaller than co-polarized components hence only a small error will be introduced. In other cases, an iterative approach can be applied.

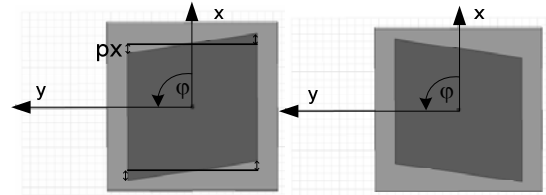

Fig. 2. Parallelogram transformation with $px$=-0.6 mm (left) and $px$=0.6 mm (right). In both cases $tx$=0 mm.

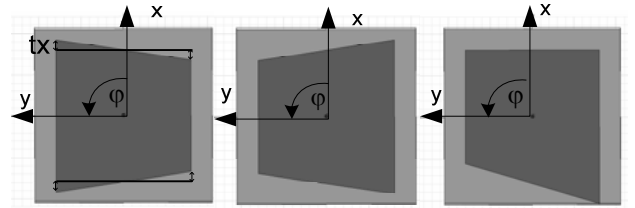

Fig. 3. Trapezoid transformation with $tx$=-0.6 mm and $px$=0 mm (left); $tx$=0.6 mm and $px$=0 mm (centre); and trapezoid-parallelogram combination with $tx$=$px$=0.6 mm (right).

By contrast, the trapezoid transformation produces phase differences between the cross-terms. It is based on tilting two opposite sides of the square patch by opposite angles, while preserving again the general original patch size. Figure 3 shows a trapezoid transformation applied to the sides perpendicular to the x-axis. Note that (9) means that cross-coupling phases can only be varied within the constraint that their average value presents a shift of 90º with respect to the average of the co-polar phases. Hence here we characterize the control of cross-coupling phases through the measure of their phase difference ($\angle(s_{12}/s_{21})$). In practical applications, there is generally at least on degree of freedom in the phases of co- or cross-coefficients which allow synthesizing the desired reflection phases.

Ansoft HFSS is used to characterize these transformations. The unit cell is based on a simple substrate stack consisting of a 4mm thick dielectric with $\varepsilon_r$=1.05 and tan$\delta$=0.00083, and perfect electrical conductors for the patch and ground plane. The operating frequency is 14.25 GHz and the lattice is set to 10.526 mm. An 8-mm wide square patch is used as a starting point in the design unless specified otherwise.

The performance of both transformations is first evaluated by independently varying the variables *px* and *tx*, which are used to measure the tilt imposed by the parallelogram and trapezoid transformations, respectively, as shown in Fig. 2-3. The magnitude of the cross-terms (note that $|s_{21}|=|s_{12}|$ as explained in section II.C) and the cross-terms phase difference ($\angle(s_{12}/s_{21})$) are plotted in Fig. 4 under oblique incidence $\varphi$=0º, $\theta$=30º. Note that the centre of the graph corresponds to the symmetric square patch (*tx*=*px*=0 mm).

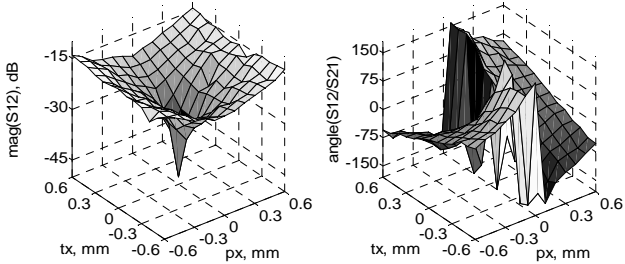

Fig. 4. Simulation results for φ =0º and θ =30º. Significant ranges of phase and magnitude variations are achieved.

As desired, the parallelogram transformation allows controlling the cross-coupling magnitude, since Fig. 4 shows that this latter is significantly affected by *px*. Similarly, the large variation of phase difference observed when varying *tx* demonstrate the effectiveness of the trapezoid transformation. However, the transformations are not purely independent, which is reflected by the variation of the magnitude with *tx* and the change of sign of the slope of the phase difference curve *versus tx* when crossing the minimum magnitude point (*px*=0 mm). This is not a limitation to proposed method since a significant range of phase differences can be achieved over a large range of magnitudes, which, in practice, would allow synthesizing almost any required reflective cell scattering matrix. In fact, the only limitation observed is that as the magnitude is increased by the increase of $|px|$, the achievable phase difference range decreases, resulting in near-optimal solutions.

### B. Dependence on the incidence angles

The dependence of the transformations on the $\varphi$ and $\theta$ incidence angles was then studied (the results for $\varphi$ =20º and $\theta$=30º are shown in Fig. 6 as an example). It is observed that for -45°<$\varphi$<45°and 135°<$\varphi$<225° a significant range of magnitude and phase variations can be achieved. Note that, as depicted in Fig. 5, the minimum magnitude point, and thereby, the turning point in the phase difference behaviour only corresponds to the symmetric square patch for $\varphi$=0º and 180º. This means that the parallelogram transformation under general oblique incidence (i.e. $\varphi$ and $\theta$ different from 0º) can also allow *reducing* the cross-polarized magnitude with respect to the symmetrical case. Finally, the range of phase difference variation slightly increases as $\varphi$ gets close to +/-45º or +/-135º.

For 45°<$\varphi$<135° and -135°<$\varphi$<-45° the parallelogram transformation provides good magnitude control but the trapezoid only achieves very small phase variations, as shown in Fig. 6. However, the solution to this limitation is straightforward; for such incidence angles the transformations should be applied to the other patch sides, namely, the ones perpendicular to the y-axis. Therefore, by selecting the sides to perturb as a function of the incidence angle $\varphi$ allows control of the cross-coefficients for all incidence angles.

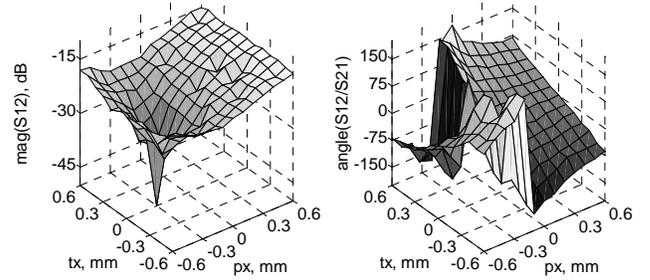

Fig. 5. Simulation results for φ =20º and θ =30º. Significant ranges of phase and magnitude variations are achieved but the minimum magnitude does not correspond to the symmetric patch.

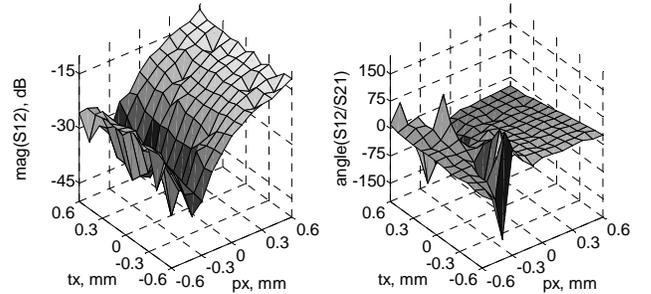

Fig.6. Simulation results for φ =75º and θ =30º. Parallelogram transformation allows controlling the magnitude but very small phase variation is achieved.

Nevertheless, a pathological case occurs under the particular 'diagonal' incidences $\varphi$ =+/-45° and $\varphi$ =+/-135°. Neither of the two transformations provides the desired control since the magnitude is only affected by the trapezoid and no significant phase difference is achieved. In fact, only a change of sign in the phase difference occurs when crossing the *tx*=0 mm 'line' as shown in Fig. 7. In this case, it was found that a small rotation combined with the aforementioned transformations allows obtaining the desired magnitude and phase difference control. Figure 8 depicts simulated results for a 6.5mm patch under $\varphi=45°$ incidence when a 5° rotation in the $+\varphi$ sense is applied to the patch. In this example, the phase difference and magnitude variation can be appreciated for $px>0$ mm.

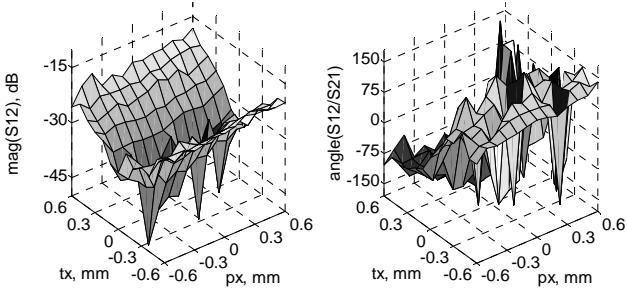

Fig. 7. Simulation results for $\varphi=45°$ and $\theta=30°$. The magnitude depends only on the trapezoid transformation and no real phase variation is achieved, only a change of sign when passing through $tx=0$mm.

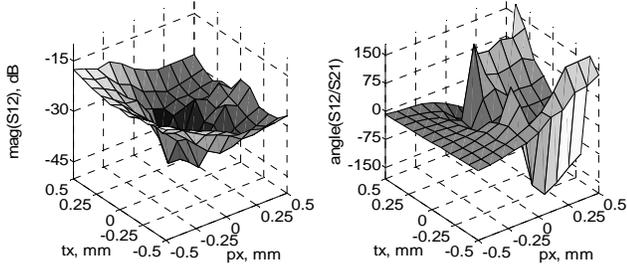

Fig. 8. Rotation of 5° combined with parallelogram and trapezoid transformations for $\varphi=45°$. Significant phase variations are achieved for $px>0$mm.

Regarding the elevation angle $\theta$, the phase difference variations became smaller as $\theta$ decrease. This is not a limitation of the trapezoid transformation but a fundamental constraint imposed by reciprocity (since at the limit [$\theta=0, \varphi_0$] and [$\theta=0, \varphi_0+\pi$] actually correspond to the same direction in space so the usual 'reciprocity condition' $s_{12}=s_{21}$ applies) and thus corresponding numerical results are omitted here for space considerations.

IV. SYNTHESIS PROCEDURE GUIDELINES

Based on this detailed analysis the following procedure for the synthesis of each reflective cell is proposed:
1) Based on the required $s_{11}$ and $s_{22}$ phases determine the patch size.
2) According to the results obtained in section III.B: (i) apply the parallelogram and trapezoid transformations to the sides perpendicular to the x-axis for incidences $-45°<\varphi<45°$ or $135°<\varphi<225°$; (ii) apply the transformations to the sides perpendicular to the y-axis for $45°<\varphi<135°$ or $-135°<\varphi<-45°$; (iii) apply a 5° rotation before the other two transformations in the pathological cases $\varphi=+/-45°$ and $+/-135°$.

The following steps describe how to apply both transformations (i.e. how to choose $px$ and $tx$) in order to achieve the targeted cross-coefficients. This discussion corresponds to the case $-45°<\varphi<45°$, but can be translated to $135°<\varphi<225°$ by reciprocity using (3).
3) Determination of $px$: (i) for $0<\varphi<45°$ use $px>0$ if an increase of the cross-terms magnitude is needed, or $px<0$ if a reduction is required, (ii) for $-45°<\varphi<0°$ do the reverse; (iii) for $\varphi=0°$ the minimum magnitude point corresponds to the symmetric case thus either positive or negative values can be used to increase the magnitude.
4) Determination of $tx$: as explained in section III.B, the value of $px$ corresponding to the minimum magnitude represents a turning point in the phase variation graph. Therefore, if $px$ larger than this turning point has been selected in step 3, an increase of the phase of $s_{12}$ requires $tx>0$, and vice-versa. By contrast, if $px$ lower than the turning point is selected, an increase of the phase of $s_{12}$ requires $tx<0$, and vice-versa. This procedure is illustrated in Fig. 9, which represents the cross-terms phases versus $tx$ for $\varphi=20°$ and $\theta=30°$. Note that two different $px$ values are shown: one larger than the one corresponding to the minimum magnitude point (i.e. $px=-0.3$mm in Fig. 6) and the other smaller.

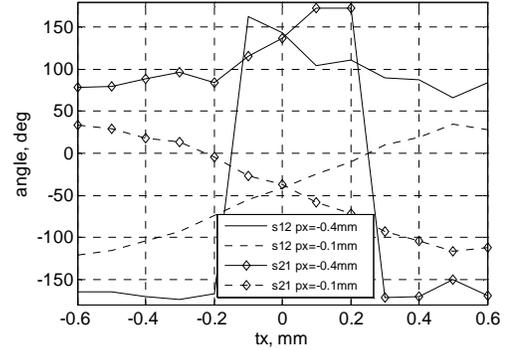

Fig. 9. Cross-terms phases for $\varphi=20°$ and $\theta=30°$